# Searching for topological semi-complete bandgap in elastic truss lattices

Yiran Hao[1,*], Dong Liu[2,*,†], Liyou Luo[2,*], Jialu Mu[3,*], Hanyu Wang[1], Zibo Liu[4], Jensen Li[2,†], Zhihong Zhu[1,†], Qinghua Guo[3,†] and Biao Yang[1,†]

[1]College of Advanced Interdisciplinary Studies, National University of Defense Technology, Changsha 410073, China.
[2]Department of Physics, Hong Kong University of Science and Technology, Clear Water Bay, Hong Kong, China.
[3]School of Physics and Electronics, Hunan University, Changsha 410082, China.
[4]State Key Laboratory of Tribology in Advanced Equipment (SKLT), Department of Mechanical Engineering, Tsinghua University, 100084, Beijing, China.
*These authors contributed equally to this work.
†Contact author: dliuan@connect.ust.hk, zzhwcx@163.com, jensenli@ust.hk, guoqh@hnu.edu.cn, yangbiaocam@nudt.edu.cn;

**ABSTRACT**. Gapless topological phases have attracted significant interest across both quantum and classical systems owing to their novel physics and promising applications. However, the search for ideal gapless topological nodes inside a clear bandgap is still lacking in elastic systems. The degenerate points are always hidden in the trivial bulk bands due to the intricate elastic modes involved. Here, we find a topological semi-complete bandgap in a three-dimensional elastic truss lattice by tuning a supporting rod, which exhibits a complete bandgap except for the inevitable topological degenerate points. Furthermore, we experimentally map the topological semi-complete bandgap and the inside nontrivial surface state arcs with a scanning laser vibrometer. The introduced scheme provides a systematic approach for the idealization of semi-complete bandgaps and thus may significantly advance the practical utility of topological phases in mechanical engineering domains.

The mathematical concept of topology has been identified as a fundamental principle underlying physical systems, particularly since the discovery of the integer quantum Hall effect [1-5], encompassing both quantum and classical realms, which is attributable to their shared mathematical framework. Extensive research has been conducted to investigate classical topological states, particularly within the realms of photonics [6-11] and acoustics [12-18], whereas their exploration in elastic waves [19-25], particularly in three dimensions, remains comparatively limited. The scalar nature of the acoustic wave allows for a simplified design analogous to the s-orbital tight-binding configuration [26-31]. Whereas an electromagnetic wave is vectorial, within a three-dimensional dielectric crystal, only two polarization states (exclusively transverse components) are preserved. However, elastic waves exhibit a more complicated behavior, involving all three vibration modes (two transverse and one longitudinal) across a three-dimensional domain. A notable consequence is that a longitudinal (compressional) mode, if it remains decoupled, will close the bandgap that might exist in the transverse (shear) modes. In contrast, when mode coupling is present, it can produce unpredictable hybridizations and complex distribution of energy bands. These factors make it difficult to isolate topological degenerate points in elasticity—they tend to be "masked" by other trivial bands at the bandgap. Several studies on finding gapless topological nodes in two-dimensional elastic systems have been conducted. Matlack et al. [32] introduced perturbative metamaterials where weakly coupled sub-units realize Dirac cone like dispersion features predicted by discrete models. Dal Poggetto et al. [33] applied optimization algorithms in spiderweb-inspired lattices to isolate Dirac-like crossings and tailor dispersion for vibration control. However, research in three-dimensional elastic systems remains scarce due to their intrinsic complexity. Developing design principles to navigate this complexity is an ongoing frontier in three-dimensional topological mechanics.

While symmetry uses the little group irreducible representation to predict band degeneracies at high-symmetry momenta [34-37], the global characteristics of the band structure are often not as thoroughly examined. Generally, the idealization of band dispersion poses significant challenges due to the unpredictable distribution of band degeneracies and bandgaps. While certain space group symmetries can enforce specific types of band connectivity, such as hourglass semimetals [38-45], most instances of band connectivity rely on various geometric details that involve infinite degrees of freedom.

Although numerous efforts have been made to enhance global predictions, such as elementary band representation [46-47], it remains challenging to develop a general design framework that adequately captures the complexity of the elastic wave band structure. On the other hand, practical implementation necessitates an ideal band structure, where the degenerate points are not masked by trivial bands, thus facilitating the observation and utilization of the associated topological properties, such as anomalous scattering, chiral anomalies, and one-way propagation in elasticity.

In this letter, the concept of topological semi-complete bandgap, i.e. the complete bandgap except for the inevitable topological degenerate points, is proposed to characterize the paradoxical situation of the coexistence of both band degeneracies and bandgaps. Although band idealization in elastic systems is notably more challenging than in other systems owing to the complex degrees of freedom involved, our study focuses exclusively on the resonance physics of a one-dimensional rod, as schematically illustrated in Fig. 1(a), to identify the topological semi-complete bandgap.

Very recently, it was shown that global band connectivity is determined by the global Hilbert band complex that comprises multi-partite graphs [34-37]. Each Hilbert band complex between neighboring momenta exhibits multiple bands connecting the degeneracies and remains connected regardless of energy permutations; thus, it is referred to as irreducible. Achieving an ideal scenario with a wide bandgap necessitates considering the global Hilbert band complex constructed between any two high-symmetry points. Figure 1(b) shows a potential configuration for the global Hilbert band complex of space group No. 198, which may reveal an obvious topological semi-complete bandgap as highlighted. Such "ideal" topological phase would allow the observation of topological surface states, free from interference by trivial bulk excitations. However, the ordering of these representations in energy is always difficult to predict, especially for high-energy bands, which are far from the simple assumption of interatomic couplings. While Weyl phonon dynamics in crystalline systems are well-established [48-49], our study proposes a structural modulation strategy for elastic rods to engineer tunable three-dimensional topological metamaterials (e.g. Fig. 1(a)), aiming to realize desired topological Weyl band structure.

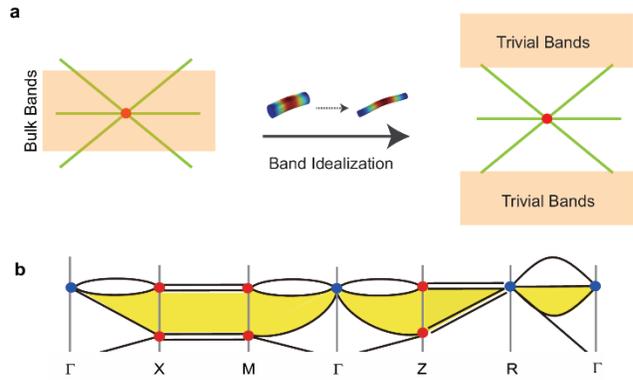

FIG. 1. Concept of a semi-complete bandgap in an elastic system. (a) Schematic illustration of band structure idealization by tuning the 1D rod. (b) Band connectivity with global Hilbert band complex for space group 198. Blue and red dots indicate symmetry-enforced degeneracies. The yellow region indicates a semi-complete bandgap around the gapless points (blue dots).

We begin by studying the diamond truss lattice, a structure that is highly prevalent in the field of topological physics and serves as a natural extension of two-dimensional mechanical graphene into three dimensions [50-53]. The mechanical properties of diamond truss lattices can be precisely tailored by modulating the geometry and connectivity of the truss elements. This design flexibility significantly enhances the ability to tailor various elastic wave properties to meet specific requirements.

Our aim is to forge a conceptual connection between this lattice and the well-studied chiral space group No. 198. For a diamond conventional unit cell, each tetrapod features four equivalent arms, as shown in Fig. 2(a). Figure 2(d) shows the corresponding band spectra, where are a six-fold degeneracy at the $\Gamma$ point and an eight-fold degeneracy at the $R$ point. One sees that the longitudinal mode (dashed lines) is fully decoupled from those transverse modes (solid lines) and straightly crosses the bandgap of the transverse modes. Consequently, there is no complete bandgap as mentioned above.

We subsequently apply a slight elongation to the arm along the (111) axes, as highlighted in pink in Fig. 2(a), while maintaining the $C_3$ symmetry. As a result, the unit cell transforms into a simple cubic crystal with space group P2$_1$3 (No. 198). We further refine the band structure by gradually reducing the radius of the pink rod and obtain the optimized design as shown in Fig. 2(b). The corresponding semi-complete bandgap is highlighted in yellow as shown in Fig. 2(e), where one sees that a spin-1 Weyl point and a charge-2 Dirac point are the inevitable degenerate points. More details regarding the evolution of the elastic truss lattice can be found in Fig. S1 (see Supplementary Material I).

To visualize the nature of the topological degeneracies, Fig. 2(f) depicts the eigenmode displacements at the spin-1 Weyl point of the final design. Each of the three degenerate modes is characterized by vibrations largely localized on the thin pink rods. This confirms that the local resonance of the thin rod is a key player: the triply degenerate point can be thought of as arising from the coupling of one local resonant mode on the thin rod with extended modes of the surrounding lattice. Because the mode energy is mostly confined in the thin rods, the frequency of this triple-point is tunable by adjusting that rod's properties, and the existence of the degenerated nodes is guaranteed by the lattice symmetry rather than fine-tuning.

The manipulation of the resonance behavior of the 1D rod is pivotal in our quest to identify ideal topological phases with predictable Hilbert band complex. The radius of the supporting rod in our design serves as a key parameter for controlling the coupling strength and tuning the bandgaps between different modes, thereby enabling the realization of a broad and distinct gap region that retains gapless points within. In 198 space group, a band complex may contain arbitrarily large number $N_c = 4n$ ($n \in \mathbb{N}$) of bands [35], and our design realizes such an ideal phonon spectrum with $N_c = 8$ as shown in Fig. 2(e). We also investigated the effect of further variations in the critical rod's radius to explore the robustness and possible emergence of multiple gaps. As the rod radius taken as 0.75 mm, 0.5 mm, 0.25 mm, and 0 (removed), the energy vertices corresponding to symmetry protected degeneracies are permutated at those high-symmetry points (from Fig. 3(a) to 3(d)). During the transformation, the modes of thin rods become decoupled from the rest of the lattice and form flat bands. Consequently, a second semi-complete bandgap (green region in Figs. 3(c, d)) opens at a higher frequency range. They further result in the alternation of the band connectivity with $N_c = 12$. This analysis underscores that our targeted gap is robust for a range of rod radii, but excessive weakening of the rod eventually shifts the topological features to a different part of the spectrum.

To verify the existence of the topological semimetal phase, we implement a detection system to characterize the topological states of our elastic meta-crystal (see Fig. 4(a)). Figure 4(b) shows the experimental setup, comprising a 3D-printed sample and a scanning laser vibrometer (OptoMET, Germany). The elastic meta-crystal sample consists of 16×16×16 unit cells, with a piezoelectric transducer affixed near the edge of the bottom surface to excite the surface states. To increase the precision of laser positioning, thin plastic patches are affixed to each unit cell in the bottom layer (further details are provided in Fig. S2 in Supplementary Material II).

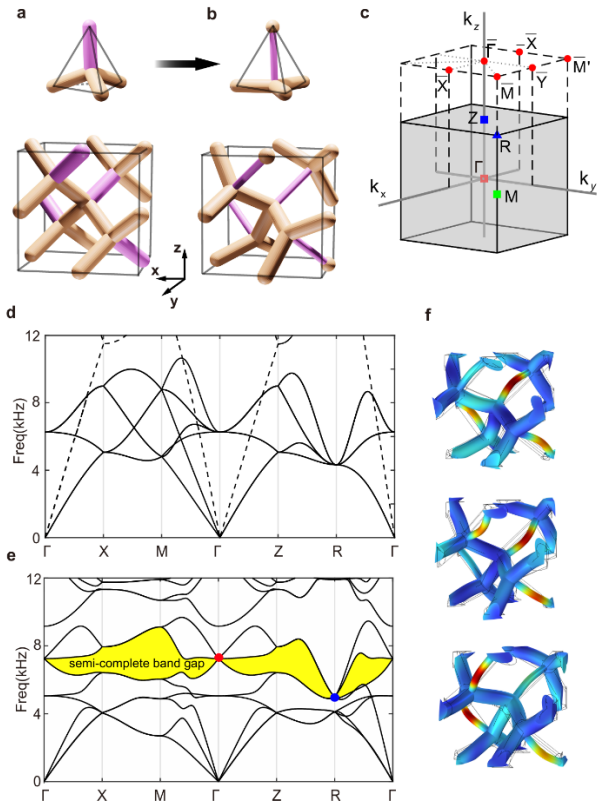

FIG. 2. Evolution of the lattice geometry and band structure. (a) Diamond initial unit cell with four equal arms. (b) The final unit cell belonging to space group P2$_1$3 (No. 198). (c) Bulk and surface first Brillion Zone with high-symmetry points labeled. (d)-(e) The corresponding band spectra of the diamond and the final decorated structure along high-symmetry path, respectively. The triple and quadruple degenerate nodes are indicated by red and blue dots, respectively. The dashed/solid lines in (d) indicate the longitudinal/transverse modes. The yellow region in (e) marks the semi-complete bandgap: the complete bandgap that is only closed by the topological gapless points. (f) Eigenmodes of the triple degeneracy at the Γ point indicated by the red dot: the three degenerate vibration patterns are plotted by highlighting the displacement magnitude on each rod (blue = low, red = high). All three modes concentrate motion in the thinner (pink) rods.

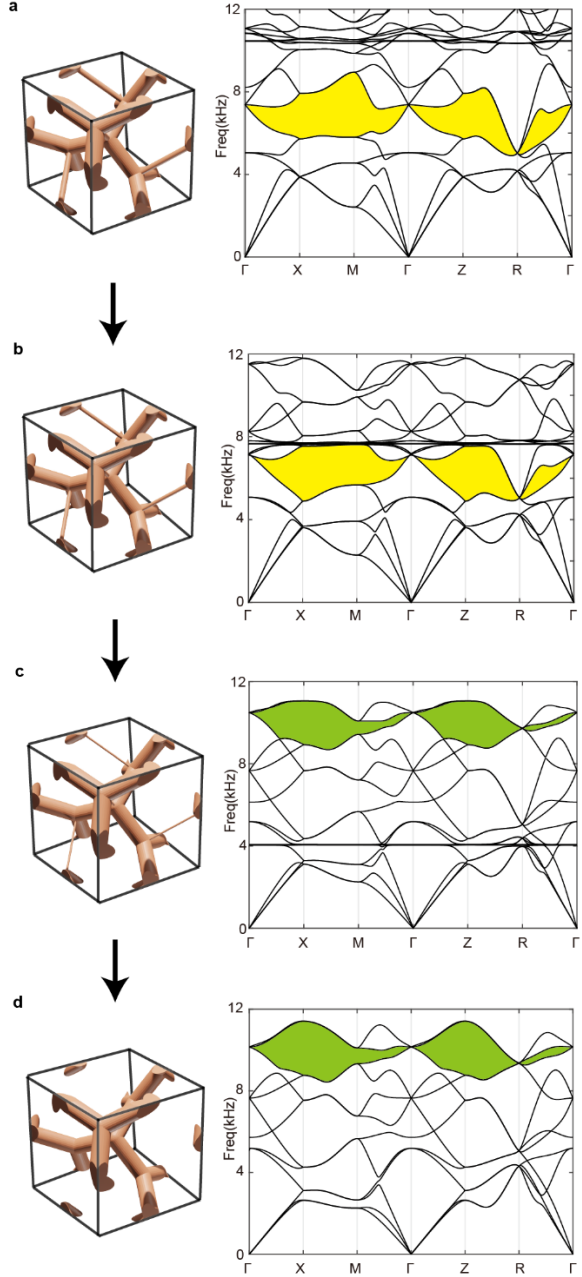

FIG. 3. Schematics of unit cells and band structures of the elastic meta-crystal. The radii of the rod along (111) direction take values of (a) 0.75 mm, (b) 0.5 mm, (c) 0.25 mm and (d) 0 mm. The yellow/green regions indicate the first/second semi-complete bandgaps.

Using the scanning laser Doppler vibrometer, we measured the out-of-plane vibration response on a 2D grid of points over the bottom surface. By performing a 2D spatial Fourier transform at fixed frequencies, we obtain the iso-frequency contour maps in the 2D surface Brillouin zone. Figure 4(c/e) shows the numerically iso-frequency contour map at frequency of 6.70/7.05 kHz for the bottom surface. In these simulations, we see clear arc-like features (curved blue lines) connecting the $\overline{\Gamma}$ point (center) to the $\overline{M}'$ point (corner) (more results of the simulated top and bottom Fermi arcs are shown in Fig. S3 in Supplementary Material III). Figure 4(d/f) shows the corresponding experimental results. Despite inevitable damping and measurement noise, the experimental data shows excellent agreement with the predicted features. We observe two prominent arcs winding around the center and the corners. In addition to the nontrivial iso-frequency arcs, the bright dots in the experimental results correspond to the contractible trivial loops in the simulations. The existence of both nontrivial and trivial bands can be further verified via theoretical/experimental results of the band spectra as shown in Fig. S4 (Supplementary Material IV).

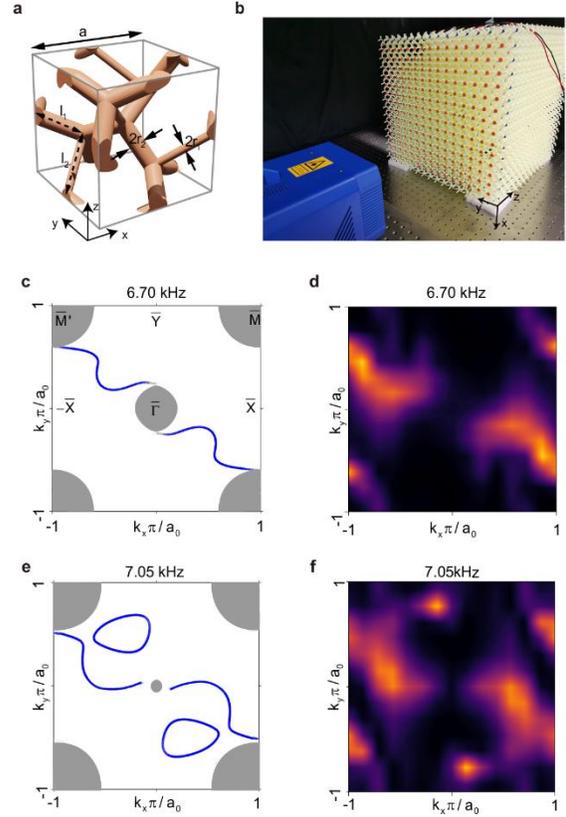

FIG. 4. Experimental demonstration of surface Fermi arcs in the elastic Weyl meta-crystal. (a) Dimensions of one unit cell of the elastic meta-crystal. The lattice constant a=30 mm, the radius of the rods $r_1$ = 1 mm and $r_2$ = 2 mm, the lengths of the rods are $l_1$ = 12.247 mm and $l_2$ = 17.321 mm. (b) Experimental set-up. The out-of-plane displacements on the bottom of the sample are measured by the scanning laser vibrometer. The sample is printed with cured photosensitive resin with the following properties: density $\rho_0$=1190 $kg/m^3$, Young's modulus $E_0 = 3.4\ GPa$, and Poisson ratio $\nu = 0.35$. (c), (e) Numerical surface state arcs at 6.70 kHz and 7.05 kHz, respectively. (d), (f) Experimentally measured bottom surface state arcs corresponding to (c) and (e), respectively.

To the best of our knowledge, this kind of "Weyl phononic crystal" is the first 3D elastic system observed to host surface Fermi-arc states within a full spectral gap. The pursuit of a broad and clean gap is important in the field of mechanical engineering, particularly in applications such as vibration isolation and energy harvesting. Given the inherent complexity of elastic frameworks, the occurrence of unpredictable mode conversions that may disrupt the desired gap is an unavoidable challenge. Consequently, topological gapless points often remain concealed within trivial bands in the three-dimensional elastic truss lattice.

In summary, by combining symmetry considerations, band connectivity analysis, and the manipulation of local resonances of 1D elastic rods, we have achieved a long-sought goal in mechanical metamaterials: a clean three-dimensional semi-complete bandgap containing topological nontrivial degenerate points. Our design framework provides a template for discovering and realizing other exotic topological phases in elasticity.